\documentclass[]{article}
\usepackage[style=bwl-FU]{biblatex}
\usepackage[utf8]{inputenc} 
\usepackage[english]{babel} 
\usepackage[T1]{fontenc} 
\usepackage{geometry}
\usepackage{fancyhdr}
\usepackage{titlesec}
\usepackage{graphicx}
\usepackage{amsmath}
\usepackage{listings}
\usepackage{geometry}
\usepackage{parskip}

\title{Exploring the stability of solar geoengineering agreements}

\author{Niklas V. Lehmann}

\date{May 2023}

\begin{document}

\maketitle
\thispagestyle{empty}
\section*{Abstract}

A simple model is introduced to study the cooperative behavior of nations regarding solar geoengineering. The results of this model are explored through numerical methods. A general result is that cooperation and coordination between nations on solar geoengineering is very much incentivized. Furthermore, the stability of solar geoengineering agreements between nations crucially depends on the perceived riskiness of solar geoengineering. If solar geoengineering is perceived as riskier, the stability of the most stable solar geoengineering agreements is reduced. However, the stability of agreements is completely independent of countries preferences.

\textbf{Keywords:} Solar geoengineering governance, Public goods games, International Agreements and Observance, Climate change

\textbf{JEL Classification:} H41, C72, Q54

\vspace{12pt}

\textbf{Funding \& Competing Interests}
 
 The author declares that no funds, grants, or other support were received during the preparation of this manuscript.

\newpage
\setcounter{section}{0}
\setcounter{page}{1}

\section{Introduction}
 
 
 Solar geoengineering is a climate engineering technique, mainly intended to mitigate the effects of global warming, by reducing the solar radiative forcing on earth. Solar geoengineering has been first proposed in the 1965 Presidential Report on the Environment (\cite{president_advisoryCommittee}).\footnote{The report outlines the possibility to brighten oceanic surfaces, which is not to be confused with stratospheric aerosol injection.} There seem to be few doubts that solar geoengineering is a technological possibility right now. In particular, stratospheric aerosol injection, depositing tiny reflective particles in the upper atmosphere, is quite well researched. How to deploy stratospheric aerosol injection is essentially public knowledge (\cite{smith2018stratosphericCost}).
 Yet the effects of different solar geoengineering methods on the biosphere are unknown. Therefore, solar geoengineering is a radical and potentially dangerous technology.

 
Aside from the danger of unforeseen ecological effects, solar geoengineering poses an enormous global public goods problem. Solar geoengineering seems to be so cheap that the global radiative forcing could be changed, to practically unlimited degree, by sovereign actors (\cite{barrett2020solar}). This raises the question: Who sets the thermostat, and how much? This is also the title of the paper by \cite{rickels2018TurningThermo} which reviews the incentives of countries to change global solar forcing. In short: many countries such as India, Brazil and Indonesia would prefer to lower solar radiative forcing. This would likely come at the expense of countries like Germany, the UK, Canada and Russia.\footnote{This is also supported by the research conducted by \cite{emmerling2017quantifying}.} Even without the context of climate change, solar geoengineering would be a highly contested international issue.
 
 Solar geoengineering has complex interdependencies with emissions abatement policy. There remains substantive uncertainty whether solar geoengineering could promote or slow emissions abatement (\cite{wagner2019moral}, \cite{WagnerRiskyGMitigation}). Existing political friction over global governance of CO2 emissions suggests that solar geoengineering could harm international relations and exacerbate conflict (\cite{Gleick1989TheIO}).\footnote{It is important to note that the pure \textit{possibility} of solar geoengineering is a sufficient condition for it to be a problem.} 
 
 In order to prevent conflict and uncautious deployment of solar geoengineering, we need to accurately forecast developments and assess how global regulation and sensible policy can foster peace and collaboration. In what follows, I try to examine how countries incentives may translate into strategies. This reasoning is agnostic to what may be desirable. However, I hope to inform the question: \textit{How might we best steer towards coordination over conflict?}
 To analyze the motives and the potentially resulting behavior around solar geoengineering, a game-theoretic model is used. Although numerous models have been used in the literature, they are either geared towards very specific questions (e.g. \cite{bas2020contesting}) or ignore fundamental options, such as the game proposed by \cite{weitzman2015voting}, which does not think of sanctions or counter-geoengineering. In this paper, I therefore simplify and analyze the 'climate-tug-of-war' model from \cite{bas2020contesting}, which results in the most simple yet sensible and \textit{extendable} model.
 The following section reviews the model in detail. Section 3 covers the results from the analysis and their implications.  
 
 \section{Model}
 
  Nations are modeled, as is usual in international relations, as unitary actors that try to minimize their losses from solar geoengineering. Nations have two  options: they can geoengineer to cool or warm the earth.\footnote{Agents are actually changing the solar forcing which is distinctly different from changing the temperature. Temperature is used throughout the entire paper only as a proxy for solar forcing, to allow for an easier read. Also, there might be nations that might want to e.g. decrease temperature, but not solar radiative forcing.} The latter refers to the possibility that countries emit more greenhouse gases, warming agents or in other ways counteract existing solar geoengineering efforts.\footnote{This could also be through conflict, sanctions or directly through technically intervening with cooling agents (see e.g. \cite{parker2018Counter}).}  Furthermore, nations can determine how strongly they want to influence the global temperature, i.e. how much they want to cool or warm the earth. It is important to emphasize that this is a real choice that policy makers (will) have. 
  Because of those options, the model is game-theoretic in nature.\footnote{The model best matches the effects of stratospheric aerosol injection as this geoengineering method is both cheap and global in effect. The effects of other methods may be more nuanced and are therefore not captured well in the analysis.}


 Nations experience a loss from their preferred temperature $y^*_i$ not being realized. Nations also experience a loss from global geoengineering activity itself.  This is due to the fact that geoengineering is risky and may lead to unforeseen ecological blowback (\cite{tang2021fate}). The assumption is that if more geoengineering is deployed, the likelihood of catastrophe rises. Geoengineering activity is expressed in units and denoted $g$. If cooling agents are deployed, the individually deployed geoengineering $g_i$ is negative. If warming is undertaken, $g_i$ is positive. The default temperature is $y=0$. The game has one period in which each nation sets its own geoengineering level $g_i$. The nations have no knowledge of the other nations geoengineering plans.  However, the nations do know about the individually preferred temperatures of other nations. The resulting equilibrium temperature is $y=\sum g$. This means that e.g., if one nation deploys one unit of a cooling agent $g_i=-1$ and the opponent two units of a warming agent $g_{-i}=2$ the result is $y=0-1+2=1$. 
 
 The loss functions $L$ of all nations are: 
 
\begin{equation}
    L_i = \underbrace{(y^*_i - y)^2}_{L\ \text{from global temperature}} + \underbrace{z(g_i^2 + g_{-i}^2)}_{L\ \text{from geoengineering}}
    \label{loss_f}
\end{equation}

 The first term of the equation is the squared difference between the global mean temperature and the individually preferred temperature, i.e. the loss from not realizing the preferred temperature. The second term of the equation describes the loss from global geoengineering activity. This term takes into account both the geoengineering undertaken by the nation itself as well as the amount of geoengineering undertaken by others.\footnote{The risk from climate engineering is a function of $g_i^2 + g_{-i}^2$ and not $\sum^{i} |g_i|$, the total amount of climate engineering activity. That is, the marginal loss from additional climate engineering is different for measures taken by the state ($g_i$) than for measures taken by others ($g_{-i}$). This may make sense if most of the risk comes from \textit{termination shock}. If multiple states engage in geoengineering, a unilateral loss of geoengineering capability is much less drastic. This assumption critically depends on whether a loss of geoengineering capability is even realistic over periods of time in which termination shock could occur.}
 The factor $z$ denotes the perceived riskiness of solar geoengineering.  If geoengineering is perceived to be very dangerous, then $z$ is high. If geoengineering is not considered risky, then $z$ converges towards zero. 
 
 What is the outcome of such a game? First off, only two-player situations are considered. This is because they are easier to analyze and may prove to be realistic enough. As stated in the introduction, there are two main camps. The nations that would benefit and those that would incur losses from reduced solar forcing. Therefore, modeling the situation as a two-player game with one nation \textit{H} preferring cooler climate $y^*_H$ and the other nation \textit{C} preferring warmer climate $y^*_C$ is not too far fetched.\footnote{For a model on coalition building see  \cite{heyen2021Coalitions}.} 
 The outcome of such a one-shot game would be that both nations (or groups of nations) deploy the following $g$:\footnote{\cite{bas2020contesting} prove this in Appendix B of their paper. The result in this paper is slightly different due to the fact that $z$ is a substitute for all occurring geoengineering cost.}
 
 \begin{align}
    g^{nc}_H &= \frac{y^*_H (z+1) - y^*_C}{z^2+2z} &
    g^{nc}_C &= \frac{y^*_C (z+1) - y^*_H}{z^2+2z}
    \label{non-cooperative_g}
\end{align}
 
 The resulting equilibrium temperature is therefore: 
 
 \begin{equation}
     y^{nc} = g^{nc}_H + g^{nc}_C = \frac{y^*_H+y^*_C}{z+2}
 \end{equation}
 
 \cite{bas2020contesting} introduce cooperation into the model by repeating the game infinitely often and allowing for grim-trigger-strategy-based cooperation. That is, the nations can make an agreement beforehand that specifies which strategy $(g^c_H,g^c_C)$ they want to play. Both nations can violate the agreement any time by playing a different $g$ than previously agreed upon. However, this terminates the agreement and both nations revert to non-cooperative behavior forever. The question therefore is: Under which conditions will nations cooperate? Nations will stick to the agreement if, and only if, cooperation has a higher expected value than non-cooperation. However, most losses come from future periods, so this critically depends upon how nations discount the future. The common discount factor with which all nations discount the future shall be denoted $\delta$. Therefore, when cooperating the losses are described by:
 
  \begin{equation}
     L^c_i + \delta * L^c_i + \delta^2 * L^c_i + \delta^3 * L^c_i + ... = \frac{L^c_i}{1-\delta} \ \ \ \text{with} \ 0 < \delta < 1
 \end{equation}
 
 Non-cooperative losses are computed in a similar fashion. If a nations breaks the agreement, this nation can set its $g^d_i$, the geoengineering deployed that deviates from the previous agreement, knowing that the other nation is going to play $g^c_{-i}$. This yields an advantage for the nation that deviates from the agreement and lowers its losses. For the individual nations, the question of whether to violate existing agreement comes down to how large non-cooperative losses plus losses from deviating are relative to losses from cooperating. In other words, does breaking the agreement yield a higher expected payoff?
 
 \begin{equation}
    \underbrace{L^d_i}_{L\ \text{from deviating}} + \underbrace{\frac{\delta*L^{nc}_i}{1-\delta}}_{L\ \text{from non-cooperation}} \geq \underbrace{\frac{L^c_i}{1-\delta}}_{L\ \text{from cooperation}}
    \label{MinDelta1}
\end{equation}

If the inequality condition in equation \ref{MinDelta1} is satisfied, then cooperation is sustainable indefinitely. Equation \ref{MinDelta1} can be rewritten as to yield the discount factor $\delta_{min}$ that is at least necessary to sustain cooperation. 

\begin{align}
    \delta &\geq \frac{L^d_i-L^c_i}{L^d_i-L^{nc}_i} \\
    \delta_{min} &= max \left[\frac{L^d_H-L^c_H}{L^d_H-L^{nc}_H};\frac{L^d_C-L^c_C}{L^d_C-L^{nc}_C}\right]
    \label{MinDelta2}
\end{align}

This factor $\delta_{min}$ is a value that encodes the incentives to break an agreement, i.e. the \textit{stability} of an agreement. If $\delta_{min}$ is high, then the nations need to value future periods highly in order to sustain cooperation. Therefore, the lower $\delta_{min}$ is, the more stable is the agreement. 


It is assumed that countries only make pareto-optimal agreements. That is, both players will not choose an agreement that could be changed in ways that would benefit both on paper. This is an important condition because it yields $g_H=g_C$. To see why, lets again look at the losses from geoengineering (equation \ref{loss_f}).

 Any possible agreement that yields a temperature $\Tilde{y}$ can be achieved with different geoengineering strategies ($g_i,g_{-i}$). That is, the amount of geoengineering that needs to be deployed to yield the global mean temperature $\Tilde{y}$ can be distributed freely across all participating nations. The second term in equation \ref{loss_f} is minimized for $g_i=g_{-i}$ or $g_H=g_C$. The risk from geoengineering with $|g|=const.$ is lowest, if the deployment is distributed across multiple nations and facilities because the risk from termination shock is assumed to be lower. There are two additional arguments for why this might happen in reality. First, it seems unlikely that nations will be willing to pass the power to engineer the climate to others.
 If, for example, only one nation would develop geoengineering capabilities ($g_i=|g|$), it would be much more tempting for that nation to break the agreement and revert to non-cooperative behavior due to the (temporary) advantage of setting the temperature unilaterally. Second, countries may want to share the cost of deployment. This argues that many nations are likely to develop the means to engineer the climate in the future. Counter-intuitively, this would incentivize cooperation and may decrease the overall use of geoengineering. 

Countries could create pareto-suboptimal agreements that are more stable. These scenarios are not considered her.

 \section{Results}
 
 
 The model can, with respect to its simplicity and abstract nature, be used to analyze the behavior of nations. The following results are in no particular order: 
 
 \textbf{Result 1: Cooperation is incentivized} 
 
 To illustrate, let $y^*_H = -2$, i.e. nations with a "hot" climate would prefer to lower temperatures two degrees from baseline. Let $y^*_C = 1$ and $z=1$. The non-cooperative Nash equilibrium is:

  \begin{align}
    g^{nc}_H &= \frac{-5}{3} &
    g^{nc}_C &= \frac{4}{3} &
    y^{nc} = g^{nc}_H + g^{nc}_C = \frac{-1}{3}
    \label{non-cooperative_example1}
\end{align}
 
 The nations that prefer lower solar radiative forcing would deploy $\frac{-5}{3}$ geoengineering units, thereby cooling the earth. The nations that prefer slightly higher solar radiative forcing would deploy warming agents ($\frac{4}{3}$ geoengineering units deployed) or would otherwise reduce the cooling effect, e.g. through sanctioning the countries that deploy cooling agents.
 Most importantly, the nations would deploy in \textit{opposite} directions, and quite remarkably so. The outcome is that the global mean temperature changes only slightly, by $\frac{-1}{3}$, in favor of the countries preferring lower temperatures. To achieve this small change, a total geoengineering effort of $|g|= \frac{5+4}{3}=3$ is deployed. Since both geoengineering and countermeasures are costly for nations, nations are incentivized to cooperate to reduce the total amount of geoengineering. By cooperating to achieve a global mean temperature of $\Tilde{y}=\frac{-1}{3}$, the nations could reduce the amount of geoengineering deployed by a factor of ten, which would benefit everyone. This is a finding consistent with other literature on the topic (see e.g.\cite{hortonMultilateralism}, \cite{heyen2019Countergeoeengineering}). This result is also consistent with standard theory in international relations. Cooperation is usually better than conflict and the latter occurs only when cooperation is not feasible for other reasons. The result is in stark contrast with early literature on free-driving and unilateral geoengineering (see e.g. \cite{weitzman2015voting}).

 
 

 
 \textbf{Result 2: The most stable agreement aims for the outcome of a non-cooperative equilibrium}
 
 
 
 The temperature $\Tilde{y}$ for which an agreement becomes most stable can be determined by minimizing $\delta_{min}$ subject to $y^*_H,y^*_C$. This can be done analytically by setting:
 
 \begin{equation}
     \frac{L^d_H-L^c_H}{L^d_H-L^{nc}_H} = \frac{L^d_C-L^c_C}{L^d_C-L^{nc}_C}
     \label{analyticalOpt}
 \end{equation}
 
 For the most stable agreement the individual incentives to break the agreement are equal.\footnote{At least in a two-player or two-alliance situation.} This is because if they weren't equal, then there would be leeway to shift the agreement in favor of the nation that is most incentivized to break the agreement on the cost of the other nation. This would increase the overall stability of the agreement, since the stability is determined by the weakest link. 
 
In this case, the most stable temperatures were determined numerically. The most stable agreement aims for the following global mean temperature:\footnote{See the code and results of the numerical analysis in Appendix A2.}
 
 \begin{equation}
     \Tilde{y} = \frac{y^*_H+y^*_C}{z+2}
 \end{equation}

 This is exactly the outcome of the non-cooperative behavior. In other words, the most stable agreement replicates the outcome of the non-cooperative situation, but reduces the costs of getting there through coordination.

\begin{figure}
    \centering
    \includegraphics[width=.75\textwidth]{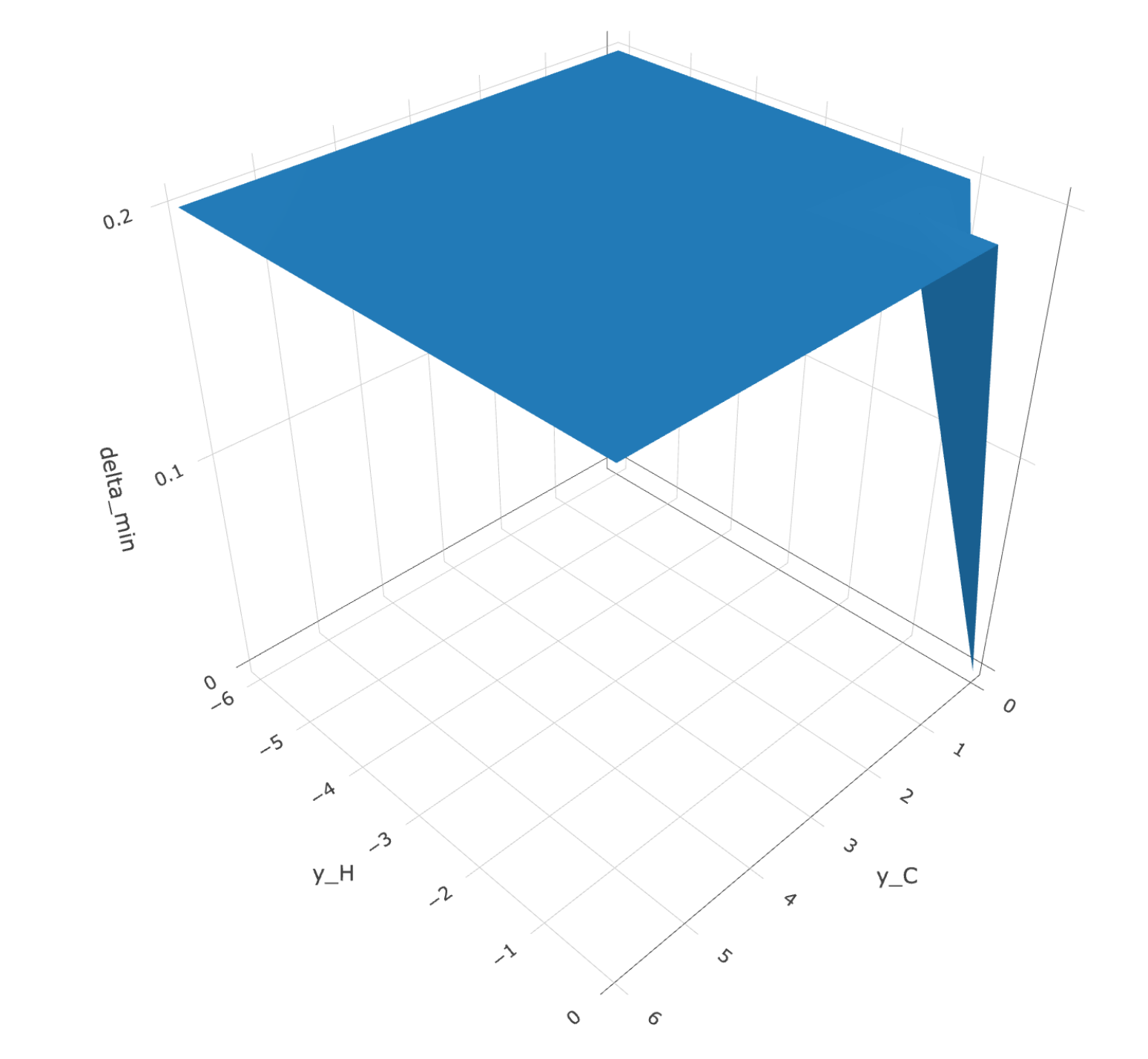}
    \caption{Stability of an agreement $\delta_{min}$, plotted against players preferences}
    \label{optDelta_z1}
\end{figure}

 What would the stability $\delta_{min}$ of such an agreement be? In figure \ref{optDelta_z1} the stability of the most stable agreement is plotted against the individual preferences $y^*_H,y^*_C$ for $z=1$.\footnote{The spike at $(0;0)$ shall be ignored. This occurs since the stability is set to zero if $y^*_H=y^*_C=0$.} The figure \ref{optDelta_z1} shows a flat plane. 
 
 \textbf{Result 3: The stability of the most stable agreements is independent of countries preferences} 
 
 In order to understand why the stability is independent of nations preferences, one needs to understand what incentivizes breaking an agreement (see equation \ref{MinDelta2}). The incentives to break a solar geoengineering agreement depend on how threatening non-cooperative behavior is ($L^{nc}_i$) and how much a nation can temporarily gain by breaking the agreement ($L^{d}_i$). Temporarily improving ones situation by breaking cooperation is the incentive and the following non-cooperative behavior is the punishment. It might seem intuitive that a nation with more extreme preferences (say $y^*_C=6$) is more incentivized to break the agreement. But there are two reasons why this is not the case. First of all, the negotiated global mean temperature $\Tilde{y}$ takes the extreme preferences already into account. A nation with extreme preferences already is partially getting its will. Second, although the global mean temperature may be far away from the extreme preferences of that nation, which incentivizes breaking the agreement, the nation is still risking unforeseen ecological blowback by pushing the temperature into extremes, which in turn reduces the incentives to break the agreement. Although the nation \textit{prefers} to have an extreme temperature, it prefers a more moderate temperature, if it has to geoengineer this change. These two effects, the higher incentive to break the treaty due to high differences between temperature preferences, and the risk of geoengineering too much, perfectly counteract each other such that preferences do not affect the overall incentives to break the agreement.
 This result holds only if $z$ is a common factor, i.e. every nation perceives geoengineering as equally risky. 
 
 
 \textbf{Result 4: The riskier solar geoengineering is believed to be, the lower is the stability of the most stable agreement} 
 
 As $z$ increases $\delta_{min}$ increases as well.\footnote{See also Appendix A2.}
 This is puzzling as the risk associated with solar geoengineering is the driving force that curtails chaotic behavior. The explanation for why the stability of agreements is reduced when geoengineering is perceived to be riskier is somewhat counter-intuitive. If solar geoengineering is perceived as riskier, i.e. $z$ is higher ceteris paribus, this has diverse effects on nations incentives. Most notably, if $z$ is higher, the non-cooperative behavior is less wasteful as nations deploy less solar geoengineering due to increased fear of unforeseen ecological effects. Ironically, this reduces the threat of non-cooperative behavior that acts as a punishment for agreement-breakers, thereby lowering the stability of agreements.
 
 This may hint at a significant problem: Currently there is quite a "silence" around solar geoengineering. Most policy-makers will be aware of the technological possibility of solar geoengineering (see e.g. \cite{EUTRACE}). Yet no country has publicly pursued a solar geoengineering program and it seems extremely unlikely that solar geoengineering will even be attempted at medium scale within the next few years. This coincides with the model results if $z$ is very large. This also makes sense since there remains massive uncertainty regarding the effects of different solar geoengineering methods. However, as research reduces this uncertainty, $z$ will become lower. Humanity may then pass through some middle-ground where solar geoengineering is perceived as very risky so that it will be hard to craft agreements that are sufficiently stable because there is little incentive to stick to them initially. This is of course a very speculative scenario, that is contingent on modeling assumptions such as $\delta$ and $z$ being shared. Yet we should take serious the possibility that the governance of solar geoengineering needs to be adapted to accommodate decreasing (or increasing) perceptions of risk.

 \section*{Conclusion and further research}
 
 The analysis of a simple game yields novel results that inform our understanding of how solar geoengineering might be deployed. However, the model ignores many important real-world aspects of international relations and more nuanced physical effects of geoengineering. Furthermore, the results for sure depend on the assumptions that have been made. Therefore, and because of the fact that solar geoengineering is ripe with politicization, we should be prepared that the social dynamics around solar geoengineering are much more chaotic than models predict. The model used in this analysis is preliminary at best and only proposed as a starting point for a more detailed analysis. There is a clear agenda for further research.

\textbf{Research path 1: Improve upon this model}
 
Enhancing the realism of the model used in this article can be achieved by incorporating features outlined by Bas and Mahajan \cite{bas2020contesting}, including non-zero and unequal deployment costs, conflict, unequal power, and imperfect monitoring. The latter seems particularly important. The incentives to break an agreement are radically altered if deviations can go unnoticed. 
 
 \textbf{Research path 2: Study other public goods games to investigate shared properties} 

It would be interesting to see which results other plausible games yield. This game could be changed with regard to the payoffs, the way nations discount the future, the number of players, how players coordinate, and many other conditions. However, tackling these questions with mathematical rigor is laborious. To accelerate research and better inform policy, I suggest leveraging recent advances in artificial intelligence. Deep neural networks have proven their capability in playing complex games like Go, making it possible to consider AI as a player in less complex games analyzed in fields such as international relations and economics (\cite{silver2016mastering}). An AI playing a game could discover Nash equilibria and cooperative strategies through self-improvement, providing valuable insights to researchers on the game's properties. Recently, DeepMind published OpenSpiel, an open source deep learning framework for games, that should be extended to public goods games (\cite{lanctot2019openspiel}). 
 
 \textbf{Research path 3: Investigate key features of other solar geoengineering methods}
 
 This model in this article draws on the idea of stratospheric aerosol injection. However, alternative methods, such as marine cloud brightening and space-based solar geoengineering, exist and offer different properties. Examining how the availability of these methods may affect behavior would be an intriguing prospect.

\textbf{Research path 4: Scenario analysis}

Conducting an experiment would provide the most compelling evidence. Subjects could play a game akin to the one described in this paper. Their behavior should be compared across various conditions, with novices, geoengineering experts, and potential policymakers included.

\vspace{24pt}

\subsection*{Acknowledgments}

I am deeply grateful for comments on early and final versions of this document by: Prof. Dr. Robert Czudaj, Prof. Dr. Gernot Wagner, Prof. Dr. Bianca Rundshagen, Vorathep Sachdev and Gideon Futerman

 \vspace{12pt}

 \newpage
 
 \printbibliography

 \newpage
 
 \section*{Appendix A1 - All functions}
 
 If the players H and C prefer $y^*_H$ and $y^*_C$ respectively, the agreed temperature $\Tilde{y}$ is achieved subject to $g_H=g_C$, then $\delta_{min}$ is a function of $y^*_H,y^*_C,\Tilde{y},z$:

\begin{equation*}
    \delta_{min_i} =  \frac{L^d_i-L^c_i}{L^d_i-L^{nc}_i}  
\end{equation*}

with:

\begin{equation}
    L^{d}_i = \left(y^*_i-\left(0,5\Tilde{y}+\frac{y^*_i - 0,5\Tilde{y}}{z+1} \right)\right)^2 + z\left((0,5\Tilde{y})^2 + \left(\frac{y^*_i - 0,5\Tilde{y}}{z+1} \right)^2 \right)
    \label{prop1_1}
\end{equation}

\begin{equation}
L^{c}_i =  (y^*_i - \Tilde{y})^2 + 0,5 z \Tilde{y}^2
\label{prop1_2}
\end{equation}

\begin{align}
        L^{nc}_i =&   \frac{(y^*_i(z^2+2z) - (y^*_H+y^*_C) (z+1) + y^*_C + y^*_H)^2}{(z^2+2z)^2} \nonumber\\
        & + \frac{z((y^*_H (z+1) - y^*_C)^2+(y^*_C (z+1) - y^*_H)^2)}{(z^2+2z)^2}
\label{prop1_3}
\end{align}
 
 \subsection*{A 1.1 Cooperative losses}
\addcontentsline{toc}{subsection}{A 1.1 Cooperative losses}

The cooperative losses are in general: 

\begin{equation*}
    L^{c}_i=  (y^*_i - \Tilde{y})^2 + z(g^c_H{}^2+g^c_C{}^2)
\end{equation*}

The agreement is based on a symmetrical deployment of $g$. 

\begin{equation*}
   g^c_C = g^c_H =  0,5\Tilde{y} 
\end{equation*}

Therefore, losses can be rewritten as:

\begin{align*}
    L^{c}_i &=  (y^*_i - \Tilde{y})^2 + z((0,5\Tilde{y})^2 + (0,5\Tilde{y})^2) \\
    L^{c}_i &=  (y^*_i - \Tilde{y})^2 + 0,5 z \Tilde{y}^2
\end{align*}

\subsection*{A 1.2 Non-cooperative losses}
\addcontentsline{toc}{subsection}{A 1.2 Non-cooperative losses}

The non-cooperative losses are in general: 

\begin{equation*}
    L^{nc}_i=  (y^*_i - (g^{nc}_H+g^{nc}_C))^2 + z((g^{nc}_H)^2+(g^{nc}_C)^2)
\end{equation*}

From \cite{bas2020contesting} Proposition 1 (p.7) or equation \ref{non-cooperative_g} in this thesis, it is known that: 

\begin{align*}
    g^{nc}_H &= \frac{y^*_H (z+1) - y^*_C}{z^2+2z} &
    g^{nc}_C &= \frac{y^*_C (z+1) - y^*_H}{z^2+2z}
\end{align*}

Inserting $g^{nc}_H$ and $g^{nc}_C$ into the loss function:

\begin{align*}
    L^{nc}_i = & \left(y^*_i - (\frac{y^*_H (z+1) - y^*_C}{z^2+2z}+\frac{y^*_C (z+1) - y^*_H}{z^2+2z})\right)^2 \\  &+ z\left((\frac{y^*_H (z+1) - y^*_C}{z^2+2z})^2+(\frac{y^*_C (z+1) - y^*_H}{z^2+2z})^2\right)
\end{align*}     

This can be rewritten as:

\begin{align*}
    L^{nc}_i = & \left(y^*_i - (\frac{y^*_H (z+1) - y^*_C + y^*_C (z+1) - y^*_H}{z^2+2z})\right)^2 \\  &+ z\left(\frac{(y^*_H (z+1) - y^*_C)^2+(y^*_C (z+1) - y^*_H)^2}{(z^2+2z)^2}\right)
\end{align*}

This can be rewritten as:

\begin{align*}
    L^{nc}_i = & \left( (\frac{y^*_i(z^2+2z) - y^*_H (z+1) + y^*_C - y^*_C (z+1) + y^*_H}{z^2+2z})\right)^2 \\  &+ \frac{z((y^*_H (z+1) - y^*_C)^2+(y^*_C (z+1) - y^*_H)^2)}{(z^2+2z)^2}
\end{align*}

This can be rewritten as:

\begin{align*}
    L^{nc}_i = &  \frac{(y^*_i(z^2+2z) - y^*_H (z+1) + y^*_C - y^*_C (z+1) + y^*_H)^2}{(z^2+2z)^2} \\ &+ \frac{z((y^*_H (z+1) - y^*_C)^2+(y^*_C (z+1) - y^*_H)^2)}{(z^2+2z)^2}
\end{align*}

This can be rewritten as:

\begin{align*}
        L^{nc}_i =&   \frac{(y^*_i(z^2+2z) - (y^*_H+y^*_C) (z+1) + y^*_C + y^*_H)^2}{(z^2+2z)^2} \\ & + \frac{z((y^*_H (z+1) - y^*_C)^2+(y^*_C (z+1) - y^*_H)^2)}{(z^2+2z)^2}
\end{align*}

\subsection*{A 1.3 Losses when breaking agreements}
\addcontentsline{toc}{subsection}{A 1.3 Losses when breaking agreements}

The individual losses in case of breaking the agreement are in general:

\begin{align*}
    L^{d}_i = (y^*_i-(\underbrace{0,5*\Tilde{y}}_{\text{opponent plays}\ g^c}+g^d_i))^2 + z\left((0,5*\Tilde{y})^2 + (g^d_i)^2 \right)
\end{align*}

The alliance that breaks the treaty will play a $g^d_i$ that minimizes its losses:

\begin{align*}
    min\ L^d_i \rightarrow \frac{\partial L^d_i}{\partial g^d_i}=0
\end{align*}

\begin{equation*}
    \frac{\partial L^d_i}{\partial g^d_i} = 2*(y^*_i-(0,5\Tilde{y}+g^d_i))*(-1)+2z*g^d_i = 0
\end{equation*}

\begin{align*}
    \frac{\partial L^d_i}{\partial g^d_i} = - y^*_i + 0,5\Tilde{y} + g^d_i + z *g^d_i = 0
\end{align*}

\begin{align*}
    g^d_i = \frac{y^*_i - 0,5\Tilde{y}}{z+1} 
\end{align*}

Inserting $g^d_i$ into the loss function yields:

\begin{align*}
    L^{d}_i = \left(y^*_i-\left(0,5\Tilde{y}+\frac{y^*_i - 0,5\Tilde{y}}{z+1} \right)\right)^2 + z\left((0,5\Tilde{y})^2 + \left(\frac{y^*_i - 0,5\Tilde{y}}{z+1} \right)^2 \right)
\end{align*}

 \newpage
 
 \section*{Appendix A2 - Numerical results and code}
 
 The goal is to find the most stable agreements given the nations preferences. 
In other terms, $\delta_{min}$ is minimized given a set of $y^*_H,y^*_C,\Tilde{y}$. What are reasonable sets of $y^*_H,y^*_C$ that should be analyzed? If $y^*_H$ and $y^*_C$  are both negative or positive, a consensus to lower or raise temperatures will form. Of interest are only cases where $y^*_H \leq 0  \And y^*_C \geq 0$. The following range of possible situations is analyzed: 

\begin{align*}
    y^*_H \in N &: 0\geq y^*_H \geq -6 \\
    y^*_C \in N &: 0 \leq y^*_C \leq 6
\end{align*}

To make the entire computation easily modifiable, all functions were replicated in R. 
The numerical analysis works as follows: The $\delta_{min}$ function is computed given the set $y^*_H = 0 \And y^*_C = 1$ so that only $\delta_{min} = f(\Tilde{y})$ remains.  Then the numerical optimization algorithm by \cite{brent1973algorithms} is used to find the $\Tilde{y}$ that minimizes $\delta_{min}$ given $z$. The result is the $\Tilde{y}$ for which the agreement is most stable. After the minimum is computed, the next set $y^*_H = -1 \And y^*_C = 1$ is computed in the same way. This goes on until all minimum points for all possible sets of preferences have been found. The set $y^*_H = 0 \And y^*_C = 0$ is obviously trivial and the value of $\delta_{min}$ is manually set to zero in this case. After the computation is finished, the results can be plotted in a three-dimensional cartesian diagram. The space between the data points is interpolated linearly so that a surface forms.

 \subsection*{A 2.1 Numerical results}
 
\begin{figure}
    \centering
    \includegraphics[width=.75\textwidth]{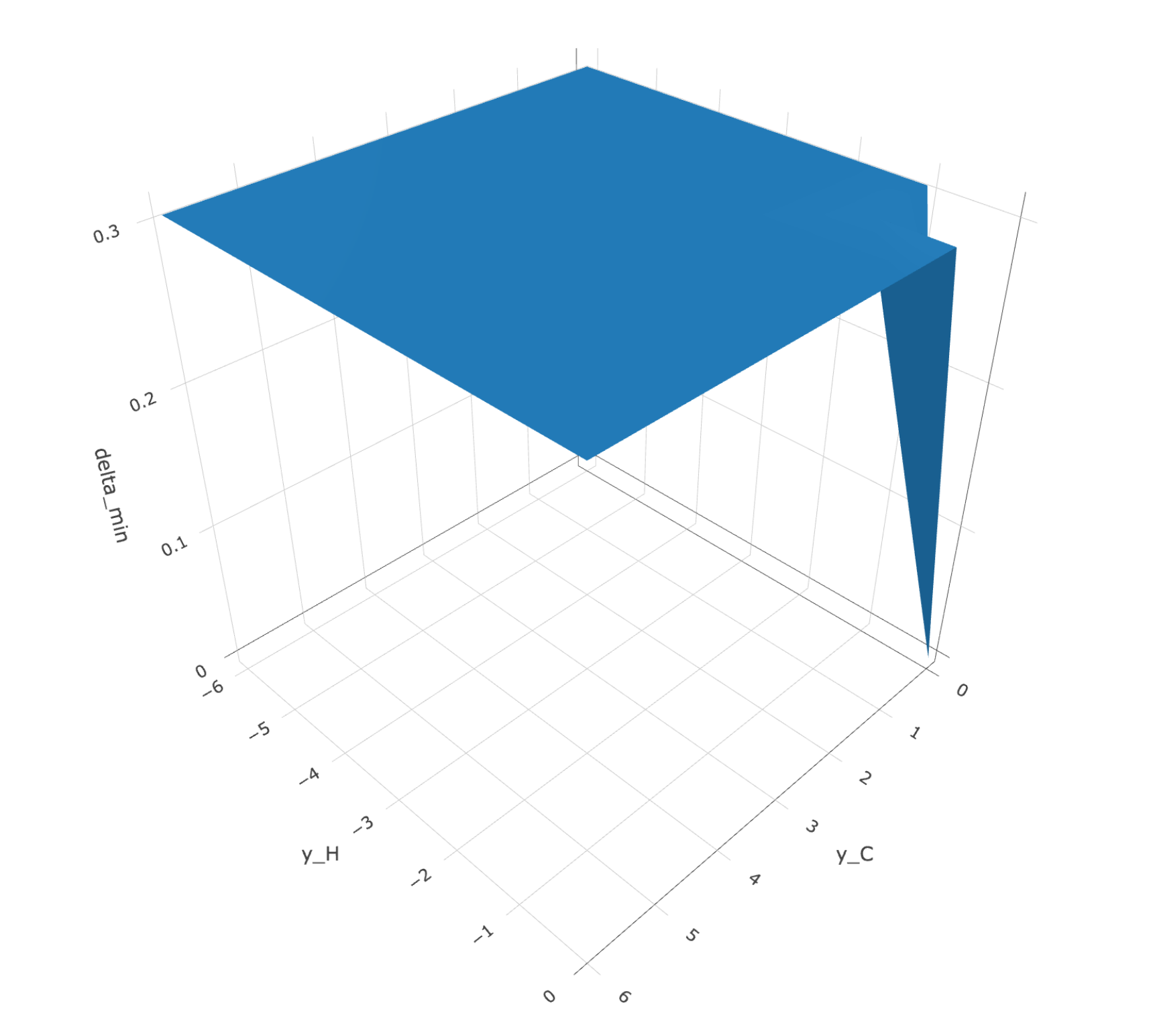}
    \caption{Stability of an agreement $\delta_{min}$ for $z=7$, plotted against players preferences}
    \label{optDelta_z7}
\end{figure}

 Plotting the results, it becomes clear that $\delta_{min}$ is only a function of $z$. This is resembled in the graphic through the fact that $\delta_{min}$ is constant across preferences (see figure \ref{optDelta_z7}). 
 
Table \ref{tab:1} shows the $\delta_{min}$ for different z. 

\begin{table}[h]
    \centering
      \caption{The effect of $z$ on $\delta_{min}$}
\begin{tabular}{|c|c|}
\hline
    $z$ &$\delta_{min}$  \\
\hline
1  & 0,2 \\
4 & 0,2857143 \\
10& 0,3125\\
\hline
\end{tabular}
    \label{tab:1}
\end{table}

\begin{figure}
    \centering
    \includegraphics[width=.75\textwidth]{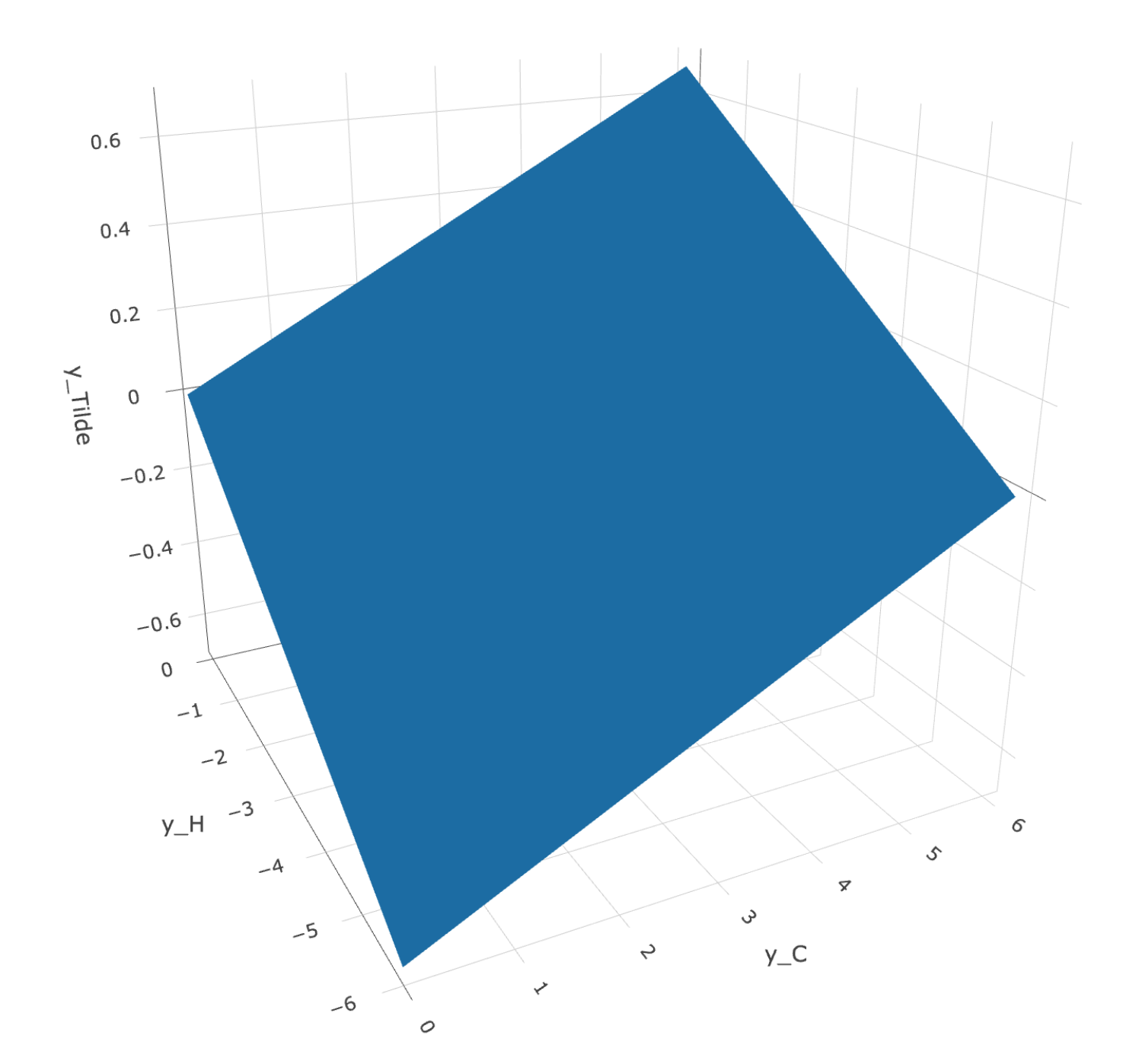}
    \caption{Mean temperature of most stable agreement $\Tilde{y}$ for $z=7$, plotted against players preferences}
    \label{optYTilde_z7}
\end{figure}

The temperatures for which the agreement is most stable is represented in a plot through a plane as well (see figure \ref{optYTilde_z7}). However, this goes to show that the most stable agreements temperature $\Tilde{y}$ is a linear function of the preferences. Experimenting with the input parameters confirms that the most stable agreements temperature is indeed the non-cooperative Nash equilibrium:

  \begin{equation}
     \Tilde{y} = \frac{y^*_H+y^*_C}{z+2}
 \end{equation}
 
 \subsection*{A 2.2 Caveats when optimizing numerically}
 
 The $\delta_{min}$ function represents the willingness of player H or C (whoever is more willing) to break the agreement. The function is $ \delta_{min} = max [\delta_{min_H},\delta_{min_C}]$. At the points where $\delta_{min_H} = \delta_{min_C}$ the $\delta_{min}$ function may inhibit a kink, if the local gradients of $\delta_{min_H}$ and $\delta_{min_C}$ are different. The $\delta_{min}$ function is therefore non-continuous which may cause trouble with the numerical optimization of the function. The results should be interpreted with caution. 

\begin{figure}
    \centering
    \includegraphics[width=.5\textwidth]{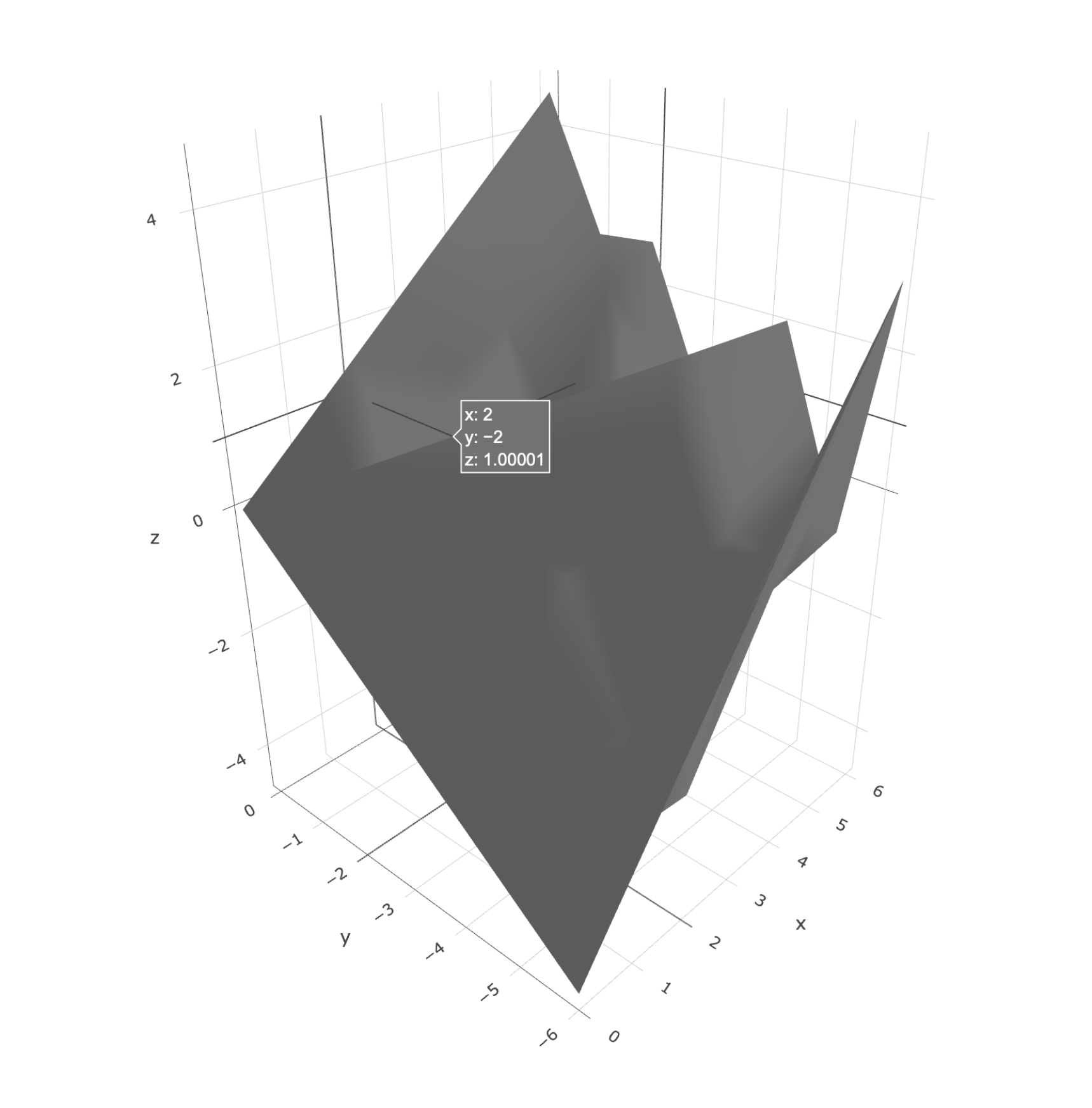}
    \caption{Temperatures for the most stable agreements form an odd shape because they are calculated incorrectly due to local singularities}
    \label{singularity}
\end{figure}

The $\delta_{min}$ function is also a rational function which further complicates the analysis. Figure \ref{singularity} shows the plot that the simulation produces given a $z=4$. The surface that forms is asymmetric, which should not occur without the possibility of conflict and $p \neq 0,5$. Some of the supposedly most stable agreements are strange. For example, the highlighted agreement shows that if $y^*_H = -2 \And y^*_C = 2$ then $\Tilde{y}=1$. The results are calculated in approximation by the computer, which is why the plot shows $\Tilde{y}$ to be 1,00001 and not 1. For $\Tilde{y}=1$ the $\delta_{min_H}$ function is undefined. At $\Tilde{y}=1$ the losses for H $L^d_H$ and $L^{nc}_H$ are equal. This is equivalent to division by zero, which causes the function $\delta_{min_H}$ to take on a value of negative infinity in the simulation. In plain terms, the agreement $\Tilde{y}=1$ is so bad for H, that it has no stability whatsoever. It is never in the interest of H to cooperate on such terms as  $L^c_H > L^d_H = L^{nc}_H$. Non-cooperation is strictly better for H than cooperation. But since $ \delta_{min} = max [\delta_{min_H},\delta_{min_C}]$ and $\delta_{min_H} = - \infty$ the value of $\delta_{min_C}$ becomes the value of $\delta_{min}$. The $\delta_{min_C}$ value of 0,0625 is very low since C is well off with the agreement. Therefore, the simulation produces the $ \delta_{min} = 0,0625$ value as the local minimum although it is not at all a sustainable agreement. Such points are in numerical simulations often called \textit{singularities} and need to be actively avoided by design. 

The algorithm was altered to incorporate a check for eligibility of potential minimum points.\footnote{This is achieved with the function in Appendix A2.4 .} That is, the range of sustainable agreements is calculated beforehand. Then, minimum points are calculated only in the possible range of reasonable agreements. For reasonable situations holds in general: 

\begin{equation}
   L^d_i < L^c_i < L^{nc}_i 
\end{equation}

Therefore the space of sustainable outcomes has its limit wherever $L^c_i$ gets as large as $L^{nc}_i$.

\begin{equation}
   L^c_i = (y^*_i -\Tilde{y})^2 + 0,5z\Tilde{y} = L^{nc}_i 
\end{equation}

This quadratic equation can be solved for $\Tilde{y}$. The limits for potential agreements are: 

\begin{equation}
   \Tilde{y}_{upper} = -0,5*\frac{-2 y^*_H}{1+0,5z} + \sqrt{\left(0,5*\frac{-2y^*_H}{1+0,5z}\right)^2 - \frac{{y^*_H}^2-L^{nc}_H}{1+0,5z}}
\end{equation}

\begin{equation}
   \Tilde{y}_{lower} = -0,5*\frac{-2 y^*_C}{1+0,5z} - \sqrt{\left(0,5*\frac{-2y^*_C}{1+0,5z}\right)^2 - \frac{{y^*_C}^2-L^{nc}_C}{1+0,5z}}
\end{equation}

The singularities are points that resemble no sustainable agreements so that they are avoided through this narrowing of the range in which the $\delta_{min}$ function is minimized.

 \newpage
 
\subsection*{A 2.3 Main function}

In the following the functions are stated as implemented in R. To allow them to fit on this page, linebreaks have been added that need to be removed before running the script. The main function is:


\begin{lstlisting}
#This function computes the most stable y_tilde 
(cooperative strategy) for each y_C,y_H (set of preferences). 

#Dependencies: NEEDS FUNCTION Calc_Boundaries pre-loaded
to avoid singularities


MostStableY_Tilde <- function(z){
	
#---VARIABLES AND DATA STRUCTURES---# 
	
	
#Dependencies: Function needs function 'delta_min' which
calculates the minimum delta_min given a 
#z is in the input 
	
y_H = 0;
y_C = 0;
y_Tilde = 0; #y_Tilde is a real number with necessary precision
temp = list(minimum = 0, objective = 0) #temporary list
a=1; #counter variable to transform matrix to vector
	
# boundaries for computation (i.e. space of sustainable outcomes)
	
boundaries = list(upper = 0, lower = 0);
	
	
#Matrix 7x7 for the optimal y_tilde
	
array_delta_min = matrix(0,7,7); 
# matrix is just an easier way to work with the data

array_y_Tilde = matrix(0,7,7); 
# matrix is just an easier way to work with the data	
	
opt_delta_min = c(1:49); 
# matrix gets converted into vector for plot

opt_y_Tilde = c(1:49); # vector for plot
	
opt = list(opt_y_Tilde=opt_y_Tilde, opt_delta_min=opt_delta_min)

	
	
#---LAYOUT COMPUTATION---# 
	
	
#Blocks of functions (decomposed parts)
	
#L^d_H = (y_H-(0.5*y_Tilde+(y_H - 0.5*y_Tilde)/(z+1) ))^2 
+ z*((0.5*y_Tilde)^2 + ((y_H - 0.5*y_Tilde)/(z+1))^2)
#L^d_C = (y_C-(0.5*y_Tilde+ (y_C - 0.5*y_Tilde)/(z+1) ))^2 
+ z*((0.5*y_Tilde)^2 + ((y_C - 0.5*y_Tilde)/(z+1))^2)
#L^nc_H = ((y_H*(z^2+2*z) - (y_H+y_C)*(z+1) + y_C + y_H)^2 
+ z*((y_H*(z+1) - y_C)^2+(y_C*(z+1) - y_H)^2))/(z^2+2*z)^2
#L^nc_C = ((y_C*(z^2+2*z) - (y_H+y_C)*(z+1) + y_C + y_H)^2 
+ z*((y_H*(z+1) - y_C)^2+(y_C*(z+1) - y_H)^2))/(z^2+2*z)^2
#L^c_H= (y_H - y_Tilde)^2 + 0.5*z* y_Tilde^2
#L^c_C= (y_C - y_Tilde)^2 + 0.5*z* y_Tilde^2
	
	
	
#---LOCAL FUNCTION delta_min = f(y_Tilde)---#
	
delta_min <- function (y_Tilde, y_C, y_H) {
	
delta_min_C = ((y_C-(0.5*y_Tilde+ (y_C - 0.5*y_Tilde)/(z+1) ))^2 
+z*((0.5*y_Tilde)^2 + ((y_C - 0.5*y_Tilde)/(z+1))^2) 
- ((y_C - y_Tilde)^2 + 0.5*z* y_Tilde^2)) /((y_C-(0.5*y_Tilde 
+ (y_C - 0.5*y_Tilde)/(z+1) ))^2 + z*((0.5*y_Tilde)^2 +
((y_C - 0.5*y_Tilde)/(z+1))^2) - (((y_C*(z^2+2*z) -
(y_H+y_C)*(z+1) + y_C + y_H)^2 + z*((y_H*(z+1) 
- y_C)^2+(y_C*(z+1) - y_H)^2))
/(z^2+2*z)^2))
	
delta_min_H = ((y_H-(0.5*y_Tilde+(y_H - 0.5*y_Tilde)/(z+1) ))^2
+ z*((0.5*y_Tilde)^2 + ((y_H - 0.5*y_Tilde)/(z+1))^2) 
- ((y_H - y_Tilde)^2 + 0.5*z* y_Tilde^2)) 
/ ((y_H-(0.5*y_Tilde+(y_H - 0.5*y_Tilde)/(z+1) ))^2 
+z*((0.5*y_Tilde)^2 + ((y_H - 0.5*y_Tilde)/(z+1))^2) 
- ((y_H*(z^2+2*z) - (y_H+y_C)*(z+1) + y_C + y_H)^2 
+ z*((y_H*(z+1) - y_C)^2+(y_C*(z+1) - y_H)^2))/(z^2+2*z)^2)
	
if(delta_min_H < delta_min_C){
		
	
return(delta_min_C)
		
		
} 
	else {return(delta_min_H)}
	
}
	




#CORE COMPUTATION
	
for (i in 0:6){ # Count down y_H 
		
		
for (j in 0:6){ #Count up y_C
			
			
#------
			
	if (j == 0 & i == 0 ){
# value can not be computed but is obviously zero
		
	#do nothing
		
	array_y_Tilde[i+1,j+1] = 0
			
	array_delta_min[i+1,j+1] = 0
			
		
	} else { # actual optimization
			
			
	#calculate range of possible agreements
								
	boundaries = Calc_Boundaries(-i,j,z);
								
	#optimization algorithm
	
	temp = optimize(delta_min, c(-i:j), y_C=j,y_H=-i,
	lower= boundaries$lower, upper = boundaries$upper);
			
	# save data to array
								
	array_y_Tilde[i+1,j+1] = temp$minimum;
			
	array_delta_min[i+1,j+1] = temp$objective;
			
		}
			
			
			
			
#------
			
		}
		
		
	}

#---RETURN VALUES FOR PLOT---#
	
	# save data to vectors
	
	
	for (m in 1:7){
	
	for (k in 1:7){
	
	opt_y_Tilde[a] = array_y_Tilde[m,k]
	opt_delta_min[a] = array_delta_min[m,k]

	a=a+1
	
	}
	
	}
	
	
	#-Return
	
	
opt =  list(    opt_y_Tilde=opt_y_Tilde,
                opt_delta_min=opt_delta_min)
return(opt)	
	
}
\end{lstlisting}

\newpage

\subsection*{A 2.4 Function to calculate boundaries}

The function that is called to calculate reasonable boundaries to optimize within is:

\begin{lstlisting}
# This function calculates the maximum range of sustainable
outcomes for a given scenario (y_H,y_C,z)



Calc_Boundaries <- function(y_H,y_C,z){
	
boundaries = list(upper = 0, lower = 0); 
#boundaries are given back in the form of a list
	
Lnc_H = ((y_H*(z^2+2*z) - (y_H+y_C)*(z+1) + y_C + y_H)^2 
+ z*((y_H*(z+1) - y_C)^2+(y_C*(z+1) - y_H)^2))/(z^2+2*z)^2; 
# should be calculated beforehand
Lnc_C = ((y_C*(z^2+2*z) - (y_H+y_C)*(z+1) + y_C + y_H)^2 
+ z*((y_H*(z+1) - y_C)^2+(y_C*(z+1) - y_H)^2))/(z^2+2*z)^2;
	
temp_upper = - (-2*y_H/(1+0.5*z))/2  
+ sqrt( (-2*y_H/(1+0.5*z)/2)^2  
- (y_H^2-Lnc_H)/(1+0.5*z)  )  ; 
	
temp_lower = - (-2*y_C/(1+0.5*z))/2 
- sqrt( (-2*y_C/(1+0.5*z)/2)^2  
- (y_C^2-Lnc_C)/(1+0.5*z)  );
	
	
boundaries$upper = 0.9999*temp_upper; 
# clean edges of the sustainable space, 
the missing piece does not matter as it is 
incredibly unlikely that the most stable outcome will be on the 
edge of the sustainable space
	
boundaries$lower = 0.9999*temp_lower;
	
	return(boundaries)
	
}
\end{lstlisting}

\end{document}